\documentclass[aps,amsmath,amssymb, twocolumn]{revtex4}
\usepackage{amssymb, amsmath}
\usepackage{bm}
\usepackage{graphicx}
\usepackage{color}
\usepackage{amsfonts}
\usepackage{latexsym}
\usepackage{subfigure}
\usepackage{lipsum}
\usepackage{cancel}

\usepackage{hyperref}

\newcommand{\unit}[1]{\ensuremath{\, \mathrm{#1}}}

\def\be{\begin{equation}}   \def\ee{\end{equation}}
\def\eq#1{{Eq.(\ref{#1})}}    \def\fig#1{{Fig.\ref{#1}}}

\begin{document}
\title{Stochastic Model of Vesicular Sorting in Cellular Organelles}
\date{\today}
\author{Quentin Vagne, Pierre Sens}
\affiliation{  Institut Curie, PSL Research University, CNRS, UMR 168, 26 rue d'Ulm, F-75005, Paris, France.}

\begin{abstract}
The proper sorting of membrane components by regulated exchange between cellular organelles is crucial to intra-cellular organisation. This process relies on the budding and fusion of transport vesicles, and should be strongly influenced by stochastic fluctuations considering the relatively small size of many organelles. We identify the perfect sorting of two membrane components initially mixed in a single compartment as a first passage process, and we show that the mean sorting time exhibits two distinct regimes as a function of the ratio of vesicle fusion to budding rates. Low ratio values lead to fast sorting, but results in a broad size distribution of sorted compartments dominated by small entities. High ratio values result in two well defined sorted compartments but is exponentially slow. Our results  suggest an optimal balance between vesicle budding and fusion for the rapid and efficient sorting of membrane components, and highlight the importance of stochastic effects for the steady-state organisation of intra-cellular compartments.

 \end{abstract}

\maketitle

One important function of membrane-bound intracellular organelles such as the Golgi apparatus or the endosome network is the sorting of membrane components secreted or ingested by the cell \cite{mellman:1996,cai:2007,jovic:2010} and their dispatch to appropriate locations via vesicular transport \cite{kelly1985pathways}. This process is regulated by molecular recognition during vesicle budding and fusion, permitted by the sensing of membrane composition by coat proteins that control vesicle budding \cite{mancias:2008,traub:2009} and by
tethers proteins and SNAREs that mediate vesicle fusion \cite{cai:2007,chen2001snare}.  The interplay between vesicle budding and fusion poses a number of interesting questions regarding the dynamics of organelles that robustly maintain distinct compositions while constantly exchanging material. Theoretical studies investigating such questions have mostly focused on steady-state, time averaged properties of dynamical compartments exchanging material \cite{Heinrich:2005,binder:2009,dmitrieff:2011,foret:2012,bressloff:2012,Ispolatov:2013}. It has been shown in particular that given sufficiently strong specificity of the budding and fusion transport mechanisms, one expects spontaneous symmetry breaking and the appearance of stable compartments with distinct compositions \cite{Heinrich:2005,dmitrieff:2011}. The inherently stochastic nature of intracellular transport  has been much less explored \cite{gong:2010,bressloff:2013}. 
Fluctuations should however be important since the typical surface area of an endosome or a Golgi cisterna ($0.2-1 \unit{\mu m^2}$) corresponds to a few tens of the  transport vesicles  (about $50-100\unit{nm}$ diameter). This explains that strong fluctuations in size and composition of early endosomes correlate with budding and fusion events~\cite{rink:2005}.

Cellular organelles are highly complex systems receiving, processing and sorting components. Here, we concentrate on one aspect of this dynamics which is the sorting of membrane components by mean of the emission and fusion of vesicles targeting specific membrane composition.  We develop a fully stochastic description of the process by which two types $A$ and $B$ of membrane components, initially mixed into a single (``mother'') compartment, can be sorted into two pure compartments containing only $A$ or $B$  components. 
The main model ingredients; selective vesicular export and homotypic fusion, are known to play an important role in both endosomes and Golgi dynamics \cite{bonifacino:2004}.
 Our model is of conceptual importance as it quantifies an important trade-off; homotypic fusion between sorted vesicles is required to form daughter compartments, but back fusion with components still in the mother compartment slows down the sorting process. This suggests the need to optimise the vesicle budding and fusion rates for efficient sorting.
Our model is also of practical interest to understand the transient response of organelles to external perturbations, such as the {\em de novo} formation of the Golgi apparatus following the redistribution of Golgi proteins to the Endoplasmic Reticulum (ER) after drug treatment \cite{glick:2002,puri:2003}.

\begin{figure}[t]
\centerline{\includegraphics[width=8.5cm]{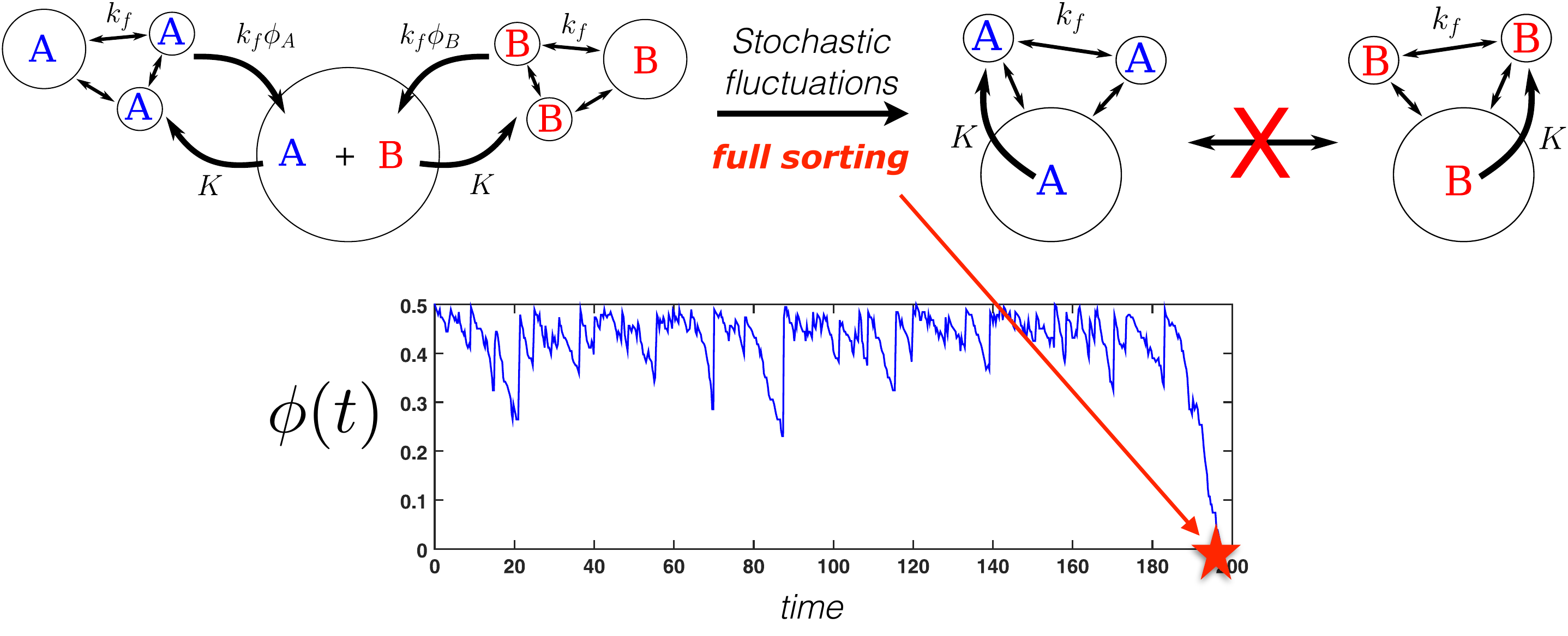}}
\caption{Sketch of the sorting mechanism. A ``mother'' compartment contains both $A$ and $B$ membrane components, which are individually exported by vesicle budding (rate $K$). Secreted vesicles with the same identity fuse with one another to create pure $A$ and $B$ compartments (rate $k_f$), and may fuse back with the mother compartment with a composition-dependent rate. Pure $A$ and $B$ compartments do not fuse together. Stochastic fluctuations lead to the permanent separation of $A$ and $B$ components into two independent distributions of pure compartments. A typical time trace for the mother compartment composition (fraction of $B$ component $\phi$) is shown. Full sorting occurs when $\phi=0$ (red star). }
\label{fig:sketch} 
\end{figure} 

.
We study the process sketched in \fig{fig:sketch}, seeking to answer two questions: (1) What is the mean first passage time to irreversible separation of $A$ and $B$ components, and (2) what is the size distribution of the pure compartments resulting from the sorting process.
 Organelles are discretised into units of area equal to that of a transport vesicle, so that a given compartment is composed of  an integer number $N$ of vesicles. We assume that each unit area has a given composition, and we consider a mixed compartment containing $N_A$ units of type $A$ and $N_B$ units of type $B$. 
The fraction of $B$ component in the compartment is written $\phi=N_B/(N_A+N_B)$. We assume that the type $B$ components are exported by vesicle budding at a rate (per site) that depends on membrane composition according to a Michaelis-Mentens scheme \cite{michaelis1913kinetik}:
\be
K_{\rm site}=Kf(\phi)\qquad f(\phi)\equiv\frac{\phi}{\phi^*+\phi}\label{budding}
\ee
where the parameter $K$ controls the vesicular emission rate and $\phi^*$ represents the selectivity of vesicle emission for the $B$ component.  We show in the Supplementary Informations (S.I.) that such kinetics may result from the selective recruitment of $B$ components by membrane-bound vesicle coat components.
The situation where both components $A$ and $B$ are exported in the same way is also studied by numerical simulations. 

Homotypic fusion between membrane compartments of similar composition controls the dynamics of both the endosomal network \cite{rink:2005} and the Golgi apparatus \cite{pfeffer:2010b}.
We implement a general homotypic fusion mechanism based on the idea that  any two compartments 1 and 2 of mixed compositions $\phi_1$ and $\phi_2$ meet at a rate $k_f$ (independent of their size), and fuse according to the probability that two random sites of their membranes are identical. This leads to the average fusion rate:
\begin{equation}
k_{\rm fusion}=k_f(\phi_{1}\phi_{2}+(1-\phi_{1})(1-\phi_{2}))\label{homotypic}
\end{equation}
This homotypic fusion scheme prevents pure $A$ and $B$ compartments ({\em i.e.} $\phi_1=1$ and $\phi_2=0$) from fusing together. Perfect and permanent sorting will thus necessarily occur at some point, after stochastic fluctuations have removed all B components from the mother compartment (see \fig{fig:sketch}). Below, we derive the mean sorting time both analytically (within some simplifying assumptions) and numerically, and we compute the steady-state size distribution of the sorted compartments.

{\bf Mean Sorting Time.}  
We consider that only the $B$ components undergo vesicular sorting, starting from a number $N_{B0}$ of type $B$ sites in the mother vesicle. In order to make analytical progress, we first assume that all $B$ components removed from the mother compartment aggregate into a single, pure $B$, daughter compartment. 
This requires to use an effective budding rate per site smaller than the actual rate ($Kf(\phi)$) to account for the back fusion of emitted vesicles with the mother compartment. The effective rate, derived in the S.I., takes the form:
$Kf'(\phi,k_f/K)=Kf(\phi)(1-\frac{\phi}{1+\phi+K f(\phi)/k_f})$. 
We compute analytically the mean first passage time (MFPT) to complete sorting of $B$ components, both in the continuous (infinite size) limit and taking finite-size effects into account. We then compare these results to numerical simulations that accurately account for the full size distribution of the sorted compartments.

\underline{\em In the large size limit:} ($N_B\gg1$), the fraction $\phi(t)$ of $B$ components in the mother vesicle may be treated as a deterministic continuous variable between fusion events. Its temporal evolution and the time $t_0$ for complete sorting of $B$ components (corresponding to $\phi(t_0)=0$) satisfy:
\begin{align}
\begin{aligned}
\frac{d\phi}{dt} &=-K(1-\phi)f'(\phi,k_f/K) \\
t_0 & =\frac{1}{K}\int_0^{\phi_0}\frac{d\phi}{(1-\phi)f'(\phi,k_f/K)}
\end{aligned}\label{diff_eq}
\end{align}

In addition to the continuous decrease of $\phi$ through vesicle emission, the compartment containing all the emitted vesicles may fuse back with the mother compartment at any instant with a probability density $k_{f}\phi(t)$. This defines a stochastic process where $\phi(t)$ continuously decreases towards zero, but is submitted to stochastic jumps that reset the system to its initial configuration $\phi_0$. Since the total number of jumps is independent of the waiting time between each jump, the MFPT to complete separation may be written
\be
\tau=t_0+\langle n_{jump} \rangle \langle t_{jump} \rangle
\label{tau_def}
\ee
where $\langle n_{jump} \rangle$ is the mean number of jump and $\langle t_{jump} \rangle$ is the mean waiting time between two jumps.

As shown in the S.I. (Eq.(S.11)) the mean sorting time can be computed exactly within our approximation, and shows two qualitatively distinct asymptotic behaviours.
When $k_f\ll K$, fusion is very unlikely and one finds $\tau\simeq t_0$ given by Eq.\ref{diff_eq}  with $f'(\phi,k_f/K)\approx f(\phi)$. In the other limit $k_f\gg K$, fusion is frequent and the mean sorting time approximates to:
\be
 \lim_{k_f\gg K}\tau=\frac{1}{k_f\phi_0}\exp\left[\frac{k_f}{K}\int_0^{\phi_0}d\phi\frac{\phi(1+\phi)}{(1-\phi)f(\phi)}\right]
\label{approx_continuous_exp}
\ee
One thus expects a transition between fast sorting and (exponentially) slow sorting when $k_f\simeq K$.

\underline{\em For small systems}, the continuous approach is not appropriate and must be replaced by the stochastic process:
\be
N_{B}\xrightarrow[{\rm budding}]{K(N_{B}+N_{A})f'(\phi,k_f/K)} N_{B}-1 \ ,\   N_{B}\xrightarrow[{\rm fusion}]{k_{f}\phi} N_{B0}
\label{evolution}
\ee
The MFPT $\tau$ may be computed exactly for this model as well (see S.I.), yielding the following asymptotic results in the limit of small and large fusion rates:
\begin{equation}
\begin{split}
& K\tau \xrightarrow[k_f/K\ll 1]{}   \sum_{N_B=1}^{N_{B0}}\frac{1}{(N_B+N_A)f(\phi)}\\
& K\tau \xrightarrow[k_f/K\gg N^2]{} \left(\frac{k_f}{K}\right)^{N_{B0}-1}\prod_{N_B=1}^{N_{B0}}\frac{\phi(1+\phi)/\phi_0}{(N_B+N_A)f(\phi)}
\end{split}
\label{tau:discrete}
\end{equation}
 with $\phi=N_B/(N_A+N_B)$. A power-law dependence $K\tau\sim (k_f/K)^{\phi_0N}$ 
 is predicted in the high fusion regime. There is however a cross-over region $1\ll k_f/K\ll N^{2}$, which is very broad for large systems, between the constant and power-law regimes of sorting time. This region corresponds to the exponential dependency $K\tau\sim e^{k_f/K}$ obtained with the continuous approximation (\eq{approx_continuous_exp}).
 
\begin{figure}[t]
\centerline{\includegraphics[width=8cm]{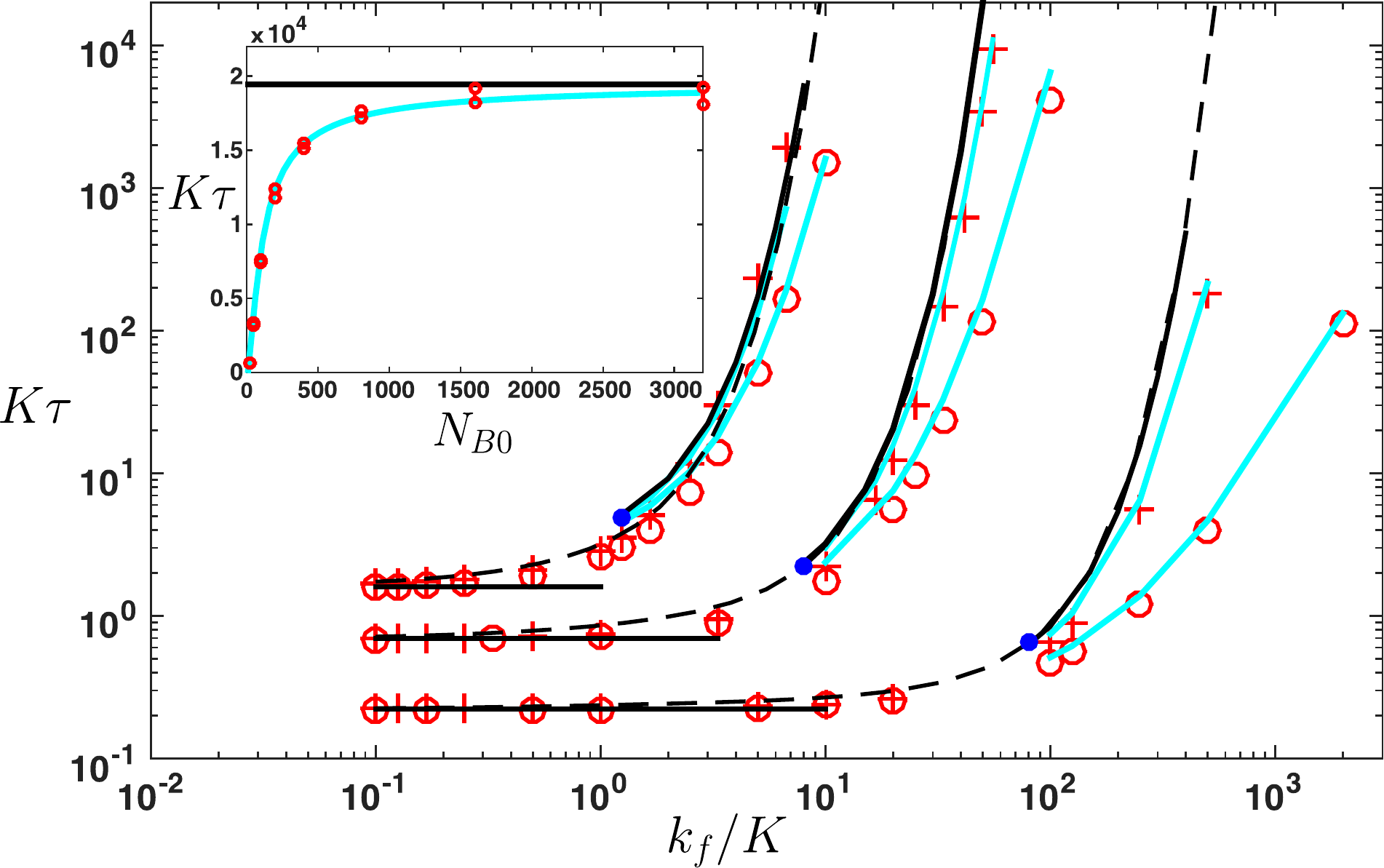}}
\caption{Dimensionless mean first passage time $K\tau$ as a function of the ratio of fusion to budding rate $k_{f}/K$ for different initial compositions and sizes. The three groups of curves correspond to different initial compositions (from left to right: $\phi_{0}=0.8$, $0.5$ and $0.2$). The simulation results are shown  for two different initial compartment sizes in each case (red circles: $N_{0}=20$, red crosses: $N_{0}=100$). The black curves show the analytical results for $N_{0}\rightarrow \infty$; dashed line is the full solution (Eq.(S.11) in S.I) and solid lines are the asymptotic limit for $K \gg k_{f}$ ($t_0$) and $K \ll k_{f}$ (\eq{approx_continuous_exp}). The light blue curves show the results of the discrete model (\eq{tau:discrete}). The crossover from fast to slow sorting (\eq{crossover}) is shown as blue dots. Inset: size dependence of the mean first passage time $K\tau$ (with $\phi_{0}=0.5$ and $k_{f}/K=50$). The error bars are standard deviations over many independent simulations.}
\label{fig:isoltime}
\end{figure}

 \underline{\em Numerical simulations} of the sorting process were performed  following a procedure described in the S.I, in the limit of strong coat selectivity ($\phi^*\ll1$ in \eq{budding}) to reduce the number of parameters. A typical time trace of the evolution of the compartment composition $\phi$ is shown on \fig{fig:sketch}. The dimensionless mean sorting time $K \tau$ (\fig{fig:isoltime}) shows excellent agreement with the analytical results and clearly exhibits the two different sorting regimes predicted analytically. Numerical studies of more realistic models, where the sorting vesicles are composed of a large number of small membrane patches and where one species may contaminate (with a low probability) the vesicles exporting the other species, are presented in the S.I. They also show  the existence of the two sorting regimes, with the same cross-over as the simpler model.

The crossover value of $k_f/K$ at which the dynamical transition between fast and slow sorting  occurs  can be obtained from mean-field arguments.
Sorting is a slow process if the system reaches a long-lived steady-state where the fusion of the sorted compartments with the mother compartment balances vesicle budding. We write a dynamical equation for the average number $N_B$ of $B$ component in the mother compartment that includes its decrease because of  vesicle budding (at a rate $K(N_A+N_B)f(\phi)$) and its increase by fusion with the $N_{B0}-N_B$ sorted components (at a rate $k_f\phi$):
\begin{equation}
\dot{N_B}=-K(N_A+N_B)f(\phi)+k_{f}\frac{N_B}{N_A+N_B}(N_{B0}-N_B)
\label{dotnb}
\end{equation} 
As discussed in the S.I., in the limit $\phi^*\ll\phi_0$ that interests us here, Eq.(\ref{dotnb}) admits a stable steady-state with $N_B\neq 0$ only if:
\begin{equation}
\frac{k_f}{K}\ge\frac{4(1-\phi_0)}{\phi_0^2}\label{crossover}
\end{equation}
This condition quantitatively predicts the cross-over between fast and slow sorting (shown by blue dots in Fig.\ref{fig:isoltime}). This result can be extended to the case where both $A$ and $B$ components can bud from the compartment. The transition from fast to slow sorting in this case is shown on \fig{fig:isoldiag}b. For an initially symmetric compartment ($N_{A0}=N_{B0}$), it occurs when $k_f/K\ge1/4$.

{\bf Size distribution of the sorted compartments}.
Although it provides an  accurate description of the different regimes of sorting dynamics and their crossover (\fig{fig:isoltime}), the assumption that all emitted vesicles gather into a single compartment is clearly an oversimplification and one expects a distribution of size for the sorted compartments.  Analytically solving the full sorting problem while accounting for this dynamically varying size distribution is very challenging. 
One may however understand the role of budding and fusion  on the compartment size distribution by investigating the steady-state size distribution of the {\em fully sorted} $B$ components, which are then isolated from the mother compartment. At the mean-field level, the number $N_n$ of pure $B$ compartments of size $n$ evolves via vesicle budding and compartment fusion according to the master equation \cite{turner:2005,foret:2012}:
\begin{eqnarray}
\frac{dN_{n}}{dt}&=&\frac{k_f}{2}\sum_{m=1}^{n-1}N_{m}N_{n-m} -k_{f}N_{n}\sum_{m=1}^{\infty}N_{m} \label{size-distrib}\\
&+&K(n+1)N_{n+1}-KnN_{n}+\delta_{n,1}K\sum_{m=1}^{\infty}N_{m}m \nonumber
\end{eqnarray}
with the constraint that the total amount of $B$ components is fixed: $\sum_{n=1}^\infty nN_n=N_{B0}$.

The  steady-state solution of \eq{size-distrib}, studied in the S.I., shows a typical power-law decay with an exponential cut-off size. Numerical simulations (\fig{fig:isoldiag}a) show the failure of the mean-field approach for fast fusion rates: $k_f/K>1$, and the appearance of a single macro-compartment containing most of the components. This can be understood qualitatively: for a compartment of size $n$, balancing the evaporation rate ($=Kn$) and the average growth rate by fusion ($\simeq k_{f}\sum_{m}mN_{m}=N_{B0}k_{f}$) yields a stationary size $n=N_{B0}k_{f}/K$, consistent with the exact solution of \eq{size-distrib} derived in the S.I. if $k_f/K\ll1$.
If $k_{f}/K \geq 1$, this predicts a compartment size larger than the total system's mass, indicating the failure of the mean-field equation \eq{size-distrib}. Assuming a macro-compartment emerges in this case, and contains a fraction $\rho$ of the  total mass $N_{B0}$,  balancing its growth by fusion with smaller compartments (flux $k_f(1-\rho)N_{B0}$) and its shrinkage by budding (flux $K\rho N_{B0}$) yields:
\begin{figure}[b]
\centerline{\includegraphics[width=9cm]{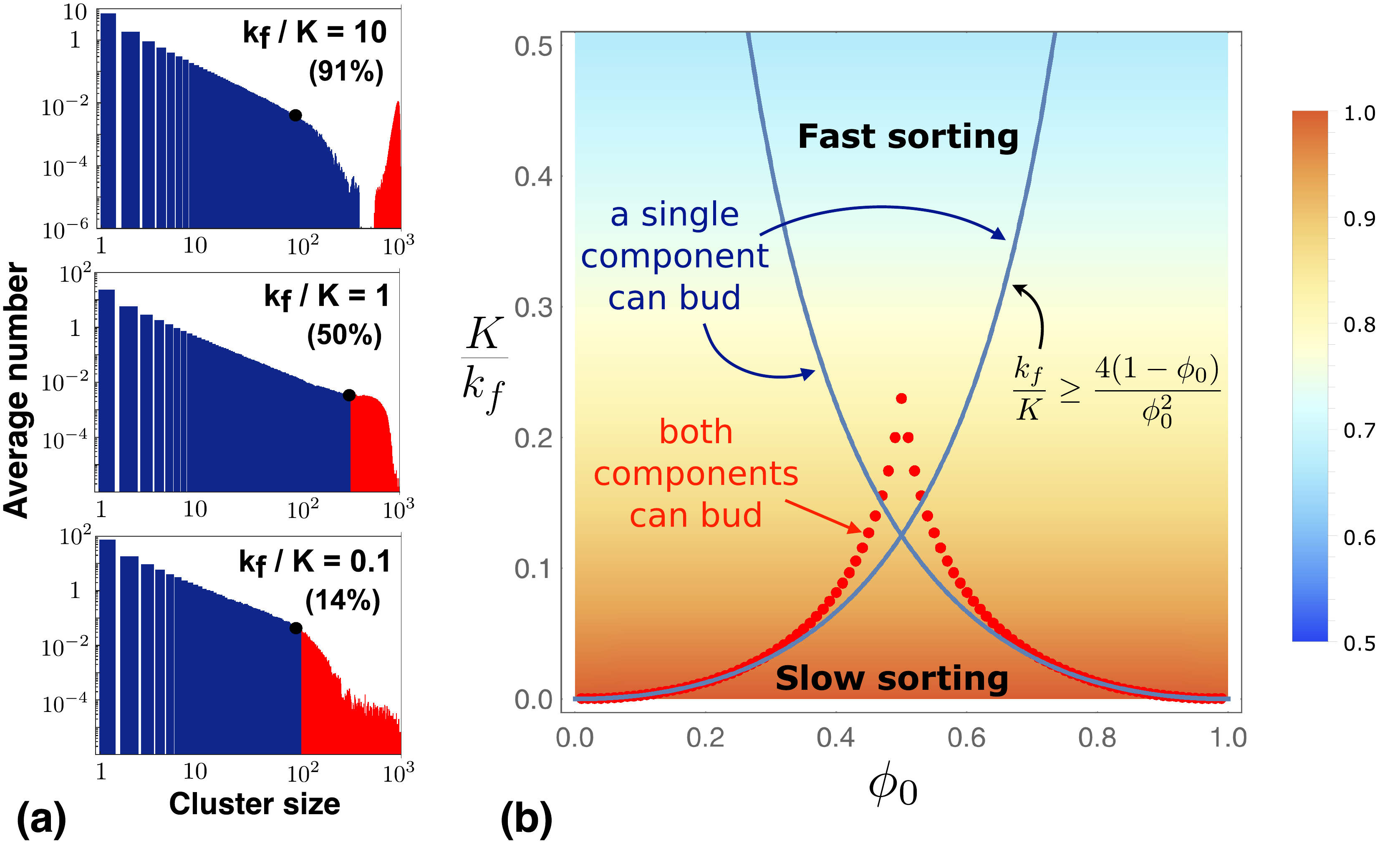}}
\caption{{\bf (a)} Steady-state size distribution of one-component compartments undergoing fusion at a rate $k_f$ and shedding vesicles at a rate $K$, for different values of the fusion to budding rate ratio $k_f/K$. The total membrane area is $N=10^3$. The single largest compartment corresponds to the red part of the distribution, which integral equals unity. It contains a fraction $\rho$ of the total area, given in parenthesis, consistent with \eq{fraction:macro}. The typical size of the  small compartments, obtained in the S.I., is displayed on each graph (black dots). {\bf (b)} Sorting phase diagram showing how the interplay between vesicular export and fusion influences both the sorting time and the distribution of sorted components. The color background represents the fraction $\rho$ of the total area contained in the largest compartment. The boundary between fast and slow sorting is indicated by the solid blue lines  when only one components can be exported in budding vesicles (\eq{crossover}), and by the red dotted line when both components can be exported. 
}
\label{fig:isoldiag} 
\end{figure} \begin{equation}
\rho=\frac{k_{f}}{k_{f}+K}.
\label{fraction:macro}
\end{equation}
Eq.(\ref{fraction:macro}) is in good quantitative agreement with the result from the full size distribution computed numerically, and shown on \fig{fig:isoldiag}a. We stress that the formation of a macro-compartment described here is a stochastic phenomenon related to finite size effects, that  it is robust upon variation of the fusion kernel in \eq{size-distrib} (see S.I.). It is thus distinct from the mean-field gelation phenomenon, which crucially depends on the size-dependency of the fusion kernel \cite{ernst:1986,leyvraz:2003}.

{\bf Discussion} The previous results show that the sorting of membrane components via selective (composition-dependent) budding and fusion events is best achieved for intermediate values of the ratio of fusion to budding rates $k_f/K$. In the absence of fusion, the composition of the mother compartment in a given component ($B$, say) removed by vesicle budding decreases with a time scale equal to the inverse budding rate. Sorting is thus very fast, but the sorted component ends up dispersed in a large number of small vesicles that are unable to fuse with one another. If fusion is allowed between membrane sites sharing the same identity, the budded vesicles are able to fuse with one another and form large compartments made solely of the sorted ($B$) components. However, fusion of the $B$ components back with the mother compartment dramatically slows down the sorting process. The resulting sorting time then strongly depends on the ratio of fusion to budding rate. The dependency is exponential for large compartments, and is a power law for small compartments (Eqs.(\ref{approx_continuous_exp},\ref{tau:discrete})).  There is thus a clear optimisation problem to solve in order to obtain two (or a few) well sorted compartments containing only $A$ or $B$ components in a relatively short time.  This is illustrated in \fig{fig:isoldiag}b, where the boundaries between fast and slow sorting regimes are shown, together with the fraction of the sorted components that are contained into a single, large compartment. 

Within this model, fast sorting  of a binary mixture of membrane components of arbitrary composition is compatible with the existence of macro-compartments containing up to $80\%$ of the sorted components. Physiologically, exchange rates between organelles vary widely, which justify our ``phase diagram'' approach to explore the full range of possible dynamical behaviour. 
Budding rates of order $K\simeq 10^{-2}/\unit{s}$ have been reported for the Golgi \cite{wang:2008}. Similar rates can be inferred from the bulk flow leaving ER exit sites \cite{thor:2009,budnik:2009}. Fusion rates are more difficult to estimate. An upper bound of  $1/\unit{s}$ is obtained by considering the time needed for a  vesicle diffusing at $1\unit{\mu m^{2}/s}$ to explore the typical Golgi dimension ($1\unit{\mu m}$). Choosing a typical ratio $K/k_f\simeq 0.03$ yields fast sorting ($\lesssim 10\unit{s}$) for $\phi_{0}<0.3$ and exponentially slow for $\phi_0>0.3$. Experimentally, {\em de novo} Golgi biogenesis after BFA-induced  Golgi redistribution into the ER appears slow ($\simeq20\unit{min}$ \cite{puri:2003}). Within our model, this corresponds to the redistributed Golgi accounting for a fraction $\phi_0=0.45$ of the ER.

While organelles along the cell trafficking pathways may to some extent be viewed as a steady-state of a complex dynamical system, specific budding and fusion events, which are at the heart of their organisation, are inherently stochastic processes. The relatively slow dynamics of {\em de novo} Golgi formation from the ER \cite{puri:2003}, as compared to the rate of ER vesiculation, suggests the existence of kinetic barriers that must be overcome stochastically. The strong fluctuations of the size and composition of early endosomes, before stochastic fluctuations lead to their full maturation into late endosomes \cite{rink:2005}, is another illustration of the need for a stochastic treatment of intra-cellular transport for physiologically relevant values of the exchange rates controlling intra-cellular organisation.  The present study proposes such a stochastic model for the process of vesicular sorting. 
Beyond its importance for identifying an optimal range of fusion to budding rate for efficient sorting, our formalism constitutes a general framework that can be used to study more complex situations of relevance to the dynamics of cellular organelles, such as the case where membrane components undergo biochemical transformation \footnote{Q. Vagne and P. Sens, {\em in\ preparation}}. 


\begin{acknowledgments}
We thank Serge Dmitrieff, Madan Rao and Matthew S. Turner for stimulating conversations.
\end{acknowledgments}

\bibliographystyle{ieeetr}

\end{document}